\documentclass[10pt,compsoc]{IEEEtran}
\pdfoutput=1 
\usepackage{subfigure}
\usepackage{natbib}
\usepackage{graphicx}
\usepackage{amsmath}
\usepackage{amssymb}
\usepackage{amsfonts}
\usepackage{epstopdf}
\usepackage{booktabs}
\usepackage{url}
\usepackage{algorithm}
\usepackage{algorithmic}
\usepackage{multicol}
\usepackage{booktabs}
\usepackage{url}
\usepackage{caption}	
\usepackage{natbib}	
\usepackage{color}

\newtheorem{definition}{Definition}[section]
\newcommand*{\RR}[1]{\textcolor{black}{#1}}

\newcommand*{\my}[1]{\textcolor{black}{#1}}	

\ifCLASSOPTIONcompsoc
\else

\fi
\ifCLASSINFOpdf
\else
\fi

\author{Rajib~Rana, Mingrui Yang, Tim Wark, Chun Tung Chou, Wen Hu
\thanks{Rajib Rana is with CSIRO, email: rajib.rana@csiro.au.}}
\begin{document}
%
\title{A Deterministic Construction of Projection matrix for Adaptive Trajectory Compression}

\markboth{IEEE Transactions on Parallel and Distributed Systems }%
{Shell \MakeLowercase{\textit{et al.}}: Bare Demo of IEEEtran.cls for Computer Society Journals}
\IEEEcompsoctitleabstractindextext{%
\begin{abstract}

Compressive Sensing\my{, which offers} exact reconstruction of sparse signal from a small number of measurements, has tremendous potential for trajectory compression. In order to optimize the compression, trajectory compression algorithms need to adapt compression ratio subject to the compressibility of the trajectory. Intuitively, the trajectory of an object moving in starlight road is more compressible compared to the trajectory of a object moving in winding roads, therefore, higher compression is achievable in the former case compared to the later. We propose an in-situ compression technique underpinning the support vector regression theory, which accurately predicts the compressibility of a trajectory given the mean speed of the object and then apply compressive sensing to adapt the compression  to the compressibility of the trajectory.


The conventional encoding and decoding process of compressive sensing uses predefined dictionary and measurement (or projection) matrix pairs. However, the selection of \my{an} optimal pair is nontrivial and exhaustive, and random selection of a pair does not guarantee the best compression performance. In this paper, we propose a deterministic and data driven construction \my{for the} projection matrix which is obtained by applying \my{singular value} decomposition to a sparsifying dictionary learned from the dataset. 

We analyze case studies of pedestrian and animal trajectory datasets including GPS trajectory data from 127 subjects. The experimental results suggest that the proposed adaptive compression algorithm, incorporating the deterministic construction of projection matrix, offers significantly better compression performance compared to the state-of-the-art alternatives. 

\end{abstract}

}

\maketitle


%
\IEEEpeerreviewmaketitle

\section{Introduction}
\label{sec:intro}
Compressive Sensing (CS)~\cite{Candes2008:Introduction}, an emerging field of information theory, has great potential for data compression. The fact that every natural signal can be sparsely represented in some sparsifying domain, is the key enabler for data compression using compressive sensing. In compressive sensing theory, a $n$ data point compressible signal $x\in\mathbb{R}^n$ can be compressed to size $m<n$, by taking $m$ projections of $x$ on to a projection matrix $\Phi \in \mathbb{R}^{m\times n}$. However, the successful recovery requires that the projection matrix $\Phi$ is incoherent to the sparsifying domain and the number of projections, $m$ is sufficient to preserve the pairwise $\ell_2$ distance of the original signal data points~\cite{bourgain1985lipschitz}. 
Typically, \my{one has $m << \footnote{much less than} n$, therefore,} compression gain is significant. 

Compressive Sensing does not offer the best compression~\cite{goyal}, however, it is particularly attractive for compression in resources improvised platforms. In addition, it is suitable for settings where there is a possibility of data-loss.  Therefore, it is ideal for embedded platforms, for example, wireless sensor nodes, and mobile phones. In particular, in our previous work~\cite{Rana:2011:AAC:1966251.1966255} we have demonstrated that compressive sensing encoding can be executed on resource constrained platform based on an 8-bit Atmel Amega 1281 microcontroller with 8 kB RAM and the reconstruction error increases gracefully with the increase of missing packets (projections). Most of the research attempts of developing compression algorithms for wireless sensor networks seek to optimize the complexity of the compression algorithms, but there is hardly any algorithm that addresses the missing data aspect of the problem. For example, S-LZW~\cite{Sadler2006}, miniLZO~\cite{miniLZo} and LEC~\cite{marcelloni2009efficient} are  three purposely adapted versions of powerful compression algorithms LZW, LZ77, and exponential-Golomb code, respectively. However, these algorithms are categorized as lossless, therefore, not robust to data loss. Even the powerful lossey algorithms adapted to wireless sensor networks , such as LTC (Light-weight Temporal Compression)~\cite{schoellhammer2004lightweight}  and Differential pulse code modulation-based optimization (DPCM-optimization)~\cite{Marcelloni20101924} are only robust to measurement error, not to the  measurement or data loss. Compressive sensing has a simplified encoding system and each projection embody the structural information of the trajectory segment, therefore, missing one measurement does not result in large error.

Intuitively, compression performance can be greatly improved if it is adapted to the movement of the object. For example,  \my{if} there is not much variation in movement, e.g., car running in motorway or animal being static, a very small number of projections should be required. On the other hand, a car moving in winding suburban roads should require large number of projections. Therefore, setting up a constant compression ratio could be either wasteful or insufficient. Compressive sensing conventionally does not offer mechanism to automatically adapt the compression.  Therefore, techniques need to be developed to adapt the compression to maximize the benefits of compressive sensing in practice. 
We propose \my{an} in-situ compression technique, which predicts the compression ratio (i.e. the number of measurements) given the speed of the moving object. 

The convention of applying the compressive sensing theory into practice is to choose a predefined sparsifying domain appropriate for the signal (e.g., the Discrete Cosine \my{Transform - DCT}) and use random projection matrices (e.g. Gaussian matrix), which are naturally incoherent with any sparsifying domain. However, given a particular dataset, it is nontrivial to find the most suitable sparsifying transform and projection matrix pair. 
\my{One of the key contributions} of this paper is the construction of the projection matrix from the dataset. We first obtain a dictionary applying  sparse coding on the trajectory datasets. We then obtain a ``special'' singular value decomposition (SVD) of the dictionary to construct the projection matrix. Since both of the projection matrix and dictionary are constructed from the dataset, they are expected to offer very good compression performance.

We case study a GPS trajectory, specifically, \my{the} animal, and pedestrian trajectory,  due to the requirement for extensive data compression. 
%
%
For example, in the CSIRO's\footnote{Commonwealth Scientific and Industrial Research Organization} virtual fencing application~\cite{Wark:2007:TAT:1262537.1262569}, animal trajectory with stimuli information \my{needs} to be sent to the base station to conform to the ethical standards. Given a 2 Hz sampling interval, from 36 cows in CSIRO's Belmont deployment, there is a requirement to transfer 7 GB of data every day over a  50 kbps line, which is not possible given available bandwidth.  Similarly, \my{in} tracking applications such as mobile forensics, child and elderly care \my{requires} the pedestrian GPS traces to be available continuously at the base stations. Given the scale of the problem, these applications require an enormous amount of data transmission within a limited bandwidth allocation\my{. Therefore},  compression is inevitable for both of these trajectories.

The key \my{contributions} of this paper can be summarized as follows:

\begin{enumerate}

\item We propose an adaptive compression framework which, underpinning the theory of compressive sensing and support vector regression, adapts the compression ratio to the compressibility of the trajecotry.

\item  We propose a data-driven and deterministic construction of projection matrix, which when combined with the trained data dictionary, offers significantly better compression compared to predefined matrix and dictionary pairs.
\item We validate the performance of the compression framework using large datasets\my{, including} pedestrian data of 91 different students and volunteers from 5 different sites, and animal data from 36 cows of CSIRO's Belmont deployment. Experimental results suggest that enabling adaptive compression with the proposed projection matrix, we can save maximum 40\% of transmission for pedestrian datasets and about 85\% transmission for cattle datasets.
\end{enumerate} 
%
%
%
%
This paper is organized as follows. In the next section (Section~\ref{sec:backGround}), we provide the background information on compressive sensing and sparse coding. In the next section (Section~\ref{sec:sparseDomain}), we describe the deterministic construction of projection matrix and the proposed adaptive compression method. We present the evaluations in section~\ref{sec:eval} and then contrast the proposed framework with the existing literature in Section~\ref{sec:related}. Finally, we conclude in Section~\ref{sec:conclusion}.

%

\section{Background}
\label{sec:backGround}
\subsection{Compressive Sensing}
\label{sec:CS}

 
\RR{We start this section by stating general conditions for compressive sensing for trajectory compression and reconstruction. In particular, we discuss sparsity, {\bf Restricted Isometry Property} and  \my{{\bf Coherence}}, which are \my{two} key criteria for compressive sensing reconstruction. We then describe the special case of adding \my{dictionaries}  into the compressive sensing framework. }
 \subsubsection{Sparsity and RIP} 
\label{sec:saprsityRip}
 For a given vector $x\in \mathbb{R}^n$, the \textit{sparsity} of $x$, namely the $\ell_0$ ``norm" of $x$, $\|x\|_0$, is defined as the number of nonzero elements in $x$. 
Compressive sensing acquires a signal $x\in \mathbb{R}^n$ by collecting $m$ linear measurements of the form $y_j = \langle a_j, x\rangle + z$, $1\leq j \leq m$, or in matrix notation $y = Ax+z$, where $A$ is a $m\times n$ measurement matrix ($m \ll n$), $z$ is the measurement error. The theory asserts that if the signal $x$ is reasonably sparse or approximately sparse, then it can be recovered, under \emph{suitable conditions} on the matrix $A$, by convex programming
 \begin{equation}\label{eqn:l1_min}
     \min_{\hat{x}\in \mathbb{R}^n}\|\hat{x}\|_1 \quad \text{subject to } \|y - A\hat{x}\|_2 \leq \epsilon,
 \end{equation}
 where $\epsilon$ is the estimate bound for $\|z\|_2$.
 
 Now we are ready to introduce one of the most important concept, the Restricted Isometry Property (RIP), which provides the \emph{sufficient condition} for signal recovery.
 
 \begin{definition}
 
     We say a matrix $A\in\mathbb{R}^{m\times n}$ satisfies the RIP with order $k$ if there exists a constant $\delta_k\in (0,1)$, such that
     \begin{equation}\label{eqn:RIP}
         (1-\delta_k)\|x\|_2^2 < \|Ax\|_2^2 < (1+\delta_k)\|x\|_2^2
     \end{equation}
     for all $x\in \Sigma_k$.
     \label{def:rip}
 \end{definition}
 
 It has been shown in \cite{Candes:08} that if the \my{RIC} $\delta_{2k}~\!\!\!<~\!\!\!\sqrt{2} -1$, then the recovered signal $\hat{x}$ from $\ell_1$~minimization \eqref{eqn:l1_min} obeys
 \begin{equation}
     \|\hat{x} - x\|_2 \leq C_0 \frac{\|x - x_k\|_1}{\sqrt{k}} + C_1\epsilon,
 \end{equation} 
 where $C_0$ and $C_1$ are constants that may only depend on $\delta_{2k}$. Several other \my{improvements on the} bounds for \my{the RIC} have also been proposed in the literature, for example, $\delta_{2k} < 0.4931$ in \cite{MoLi:11}, and $\delta_k<0.307$ in  \cite{CaiWangXu:10b}. However, the results in \cite{DaviesGribonval:09} indicate that $\frac{1}{\sqrt{2}} \approx 0.707$ is likely the upper bound of $\delta_{2k}$. 

\my{The RIP can hold for sensing matrices $A = \Phi \Psi$, where $\Psi$ is an arbitrary orthonormal basis and $\Phi$ is an $m \times n$ measurement matrix drawn randomly from a suitable distribution}, where 
\begin{eqnarray}
 m \geq Ck \log{n/k} \label{eqn:mCondition}.
\end{eqnarray}
In particular, compressive sensing theory suggests four different random matrices: 1) \my{constructing $\Phi$} by sampling i.i.d. entries from the normal distribution with mean $0$ and variance $1/m$, 2) \my{constructing $\Phi$} by sampling i.i.d. entries from a symmetric Bernoulli distribution \my{$(P(\Phi_{i, j} = \pm1/m) = 1/2)$} or other sub-gaussian distribution, 3) \my{constructing $\Phi$} by sampling $n$ column vectors uniformly at random on the unit sphere of $\mathbb{R}^m$ and 4) \my{constructing $\Phi$} by sampling a random projection $P$ as in Incoherent Sampling‚ and \my{normalizing $\Phi = \sqrt{n/ m} P$}, \my{where Incoherent Sampling referring} to choosing incoherent \my{$\Phi$} and $\Psi$ pair, i.e., no or very small correlation amongst \my{$\Phi$} and $\Psi$. For example, a randomly generated Gaussian or binary $\pm 1$  matrix is incoherent with any fixed orthonormal basis. If one fixes $\Psi$ and populates \my{$\Phi$} as in 1) and 4), then with overwhelming probability, the matrix \my{$A = \Phi \Psi$} obeys
the RIP provided that \eqref{eqn:mCondition} is satisfied, where C is some constant depending on \my{the RIC}.

One important point to note \my{is} that compressive sensing theory is based on the assumption that the signal is itself sparse, or has sparse representation in some basis or tight frame \my{$\Psi$}. However, in practice the sparsity of a signal is often not expressed in terms of an orthonormal basis or tight frame, but in terms of an overcomplete and redundant dictionary $D$. This can be confirmed from the widespread use of overcomplete dictionaries in fields like signal processing and data analysis. An overcomplete dictionary $D$ has possibly many more columns than rows. 
Candes et al. in \cite{CandesEldarNeedellRandall:11} have come up with an alternative to RIP, the restricted isometry property adapted to $D$ \my{(abbreviated D-RIP)}.

\subsubsection{\my{Coherence}}
\my{Another metric frequently used in CS is the coherence. Here we recall the definition of the mutual coherence from~\cite{Candes2008:Introduction} as an example.}

\my{The mutual coherence $\mu(A)$ for the sensing matrix $A = \Phi D$ is defined as
\begin{equation}
	\mu(A) = \max_{i < j} \frac{|\langle a_i, a_j \rangle|}{\|a_i\|_2 \|a_j\|_2},
\end{equation}
where $a_i$ and $a_j$ denote columns of $A$.
}


Generally speaking, the coherence measures the largest correlation between any two elements of $\Phi$ and $D$.  If $\Phi$ and $D$ contain \my{highly} correlated elements, the coherence is large. Otherwise, it is small. The role of the coherence is straightforward: the smaller the coherence \my{is}, the fewer measurements are needed for successful reconstruction.

\subsection{Learning Sparsifying Dictionary}
\label{sec:sparseDomain}
\label{sec:SparseCodingForTrajectoryCompression}
Given a set of vectors $X = \{\vec{x_1}, \vec{x_2}, ..., \vec{x_P}\} \in \mathbb{R}^n$, sparse coding is the technique of learning a dictionary $D$ that minimizes the loss function
\begin{eqnarray}
\ell(\vec{x},D) = \arg \min_s \frac{1}{2}||\vec{x} - \Psi \vec{s}||^2 + \lambda ||\vec{s}||_1 \label{eqn:sparse_coding}
\end{eqnarray} 
A smaller value of the loss function indicates that the vector set $X$ can be well represented by $D$.

The first part of \eqref{eqn:sparse_coding} minimizes the $\ell_2$ distance between the original and estimated vector. The second part of the equation minimizes the $\ell_1$ norm of the coefficient vector. Precisely, the second part seeks to minimize the non-zero coefficients in $s$. Sparsity of $s$ can be controlled by the regularization parameter. For example, if we use a large $\lambda$, the coefficient vector $s$ will be sparse and viceversa. However, a very large value of $\lambda$ can make the weight vector to be zero which can cause under-fitting problem. 

The value of $D$ should also not let grow too big, otherwise $s$ will become very small. The common practice is to keep the $\ell_2$ norm of each column of $D$ within 1. 

Consequently, the loss function \my{needs} to be minimized subject to both $s$ and $D$. \textcolor{black}{However, the problem of minimizing the loss function is not convex with respect to $D$.}  One way to minimize the problem with respect to $D$ is to jointly optimize~\eqref{eqn:sparse_coding} with respect to $D$ and $s$ as in~\eqref{eqn:sparse_coding_new}.
\begin{eqnarray}
\ell(x,D) = \sum_{i = 1}^P\arg \min_{s,D} (\frac{1}{2}||\vec{x_i} - D \vec{s_i}||^2 + \lambda ||\vec{s_i}||_1) \nonumber\\
s.t. ||\vec{d}_j||^2 \leq 1 \forall_j = 1,..,n	\label{eqn:sparse_coding_new}
\end{eqnarray} 

The optimization problem in~\eqref{eqn:sparse_coding_new} although \my{is} not convex with respect to $D$ and $s$ simultaneously, it is convex with respect to each of the variables $D$ and $s$, when the other one is fixed. 
Therefore, practically it can be solved in two steps: first, learning the sparse coefficients keeping the dictionary fixed, and then learning the dictionary keeping the coefficients fixed~\cite{Mairal:2009:ODL:1553374.1553463,lee07}.  

Several methods have been proposed in the literature to solve \eqref{eqn:sparse_coding_new}, out of \my{which} we consider \my{SPAMS}~\cite{Mairal:2009:ODL:1553374.1553463}, which is a recently proposed sparse coding technique. For coefficient learning, \my{SPAMS} \my{uses} the LARS-Lasso algorithm, which is a homotopy method~\cite{Osborne99anew} that provides the solutions for all possible values of $\lambda$. The key justification for choosing SPAMS is because for dictionary learning it uses block-coordinate descent with warm restarts~\cite{Bertsekas99}, which guarantees the convergence to a global optimum. Furthermore, SPAMS can process dynamic training data changing over time by processing the signals in mini-batches.

\section{Paper Contributions}
\subsection{Determinstic Construction of Projection Matrix.}
\label{sec:mTermSVD}
Recall from Section~\ref{sec:saprsityRip} that many types of random measurement matrices have a small restricted isometry constant \cite{CandesTao:06, MendelsonPajorTomczak-Jaegermann:08, RudelsonVershynin:08, BaraniukDavenportDeVoreWakin:08}. However, the random matrices are somewhat generic\my{. Therefore}, they may not always offer the smallest \my{RIC}. In fact, it remains an open question how to construct a good universal deterministic measurement matrix for accurate recovery. In this work we propose a deterministic \my{and} data driven construction of measurement matrices resulting in sensing matrices with {\bf small coherence} and {\bf RIC} for accurate recovery.

It is possibie to construct a deterministic measurement matrix based upon the representative training matrix or dictionary. For instance, in \cite{Carin:12}, the authors take the singular value decomposition (SVD) of a training matrix $D$: $D = U\Lambda V^T$, and randomly choose columns from $U$ as the rows for the measurement matrix \my{$\Phi$}. 
A singular value decomposition of matrix $D\in \mathbb{R}^{m \times n}$ is $D = U \Lambda V^{*}$, where \my{$U \in \mathbb{R}^{m \times m}$ and $V\in \mathbb{R}^{n\times n}$ are unitary matrices, and $\Lambda \in \mathbb{R}^{m \times n}$ is a rectangular matrix with nonnegative values in the diagonal and zeros elsewhere}. 
The columns of $U$ are orthonormal, therefore, {\bf uncorrelated} to the elements of the dictionary $\Psi$ in general.  Consequently, the projection matrix proposed in \cite{Carin:12} is incoherent with the dictionary \my{$D$} - which is one of the requirements \my{for} successful reconstruction.

In this paper, we propose a slightly different approach with better performance for constructing the measurement matrix. We again calculate the SVD $D = U\Lambda V^T$ of the dictionary $D$ we learn from the data. However, instead of randomly selecting $m$ columns from $U$, we only take the first $m$ columns of $U$ as the rows of the measurement matrix \my{$\Phi$}. 
The interrelationship between \my{eigenvalue} decomposition and singular value decomposition forms the basis for the better performance of our proposed method. This is because, the columns of $U$ are eigenvectors of $DD^{*}$,
therefore, by choosing the ``first'' $m$ columns of $U$, we choose the columns correspond to the largest \my{eigenvalues}. The inclusion of largest eigenvalues offer smaller Restricted Isometry Constant \my{(RIC)}, which consequently offers better reconstruction using compressive sensing. On the other hand, authors in~\cite{Carin:12}, cannot guarantee the inclusion of columns of $U$ correspond to largest Eigen values and fail to perform as good as our approach.

 \subsection{Adaptive Compression} 
\label{sec:adaptiveCompression}
\RR{Practical implementation of the proposed compression framework requires in situ prediction of number of measurements. A conservative alternative would be to use the historical maximum value, however, this could lead to poor compression performance. For example, if the subject is static, minimal number of measurements should be required, therefore, if we use the historical max, it would be enormous wastage - in particular in the wireless sensor network, where data throughput is scarce. }

\RR{Intuitively, the speed of the object is correlated to the number of measurements. For example, the vehicle trajectory of a car running in highway should require small number of  measurements compared to recovering a vehicle trajectory of a car running in winding suburban roads. In Table~\ref{tab:lookup}, we report the correlation between speed of a trajectory segment and number of measurements required to  reconstruct trajectory segments within one meter accuracy. We observe that mean speed has the strongest correlation for both pedestrian and vehicle datatsets. }

\RR{We use support vector regression to model the correlation between mean speed of the object and number of measurements. Our objective is to predict the number of projections from the mean speed of a trajectory segment.}

\RR{\begin{table}[t]
\centering
\caption{Correlation between speed and number of measurements.}
\begin{tabular}{|c|c|c|}
\hline
&\multicolumn{2}{c|}{Correlation Coefficient}\\
\cline{2-3}
Speed &pedestrian& animal\\
parameter(s)&&\\\hline
mean&0.6&0.7\\\hline
variance&0.1& 0.6\\\hline
median&0.6& 0.6\\\hline
maximum&0.1&0.6\\\hline
minimum&0.2& 0.1\\\hline
\end{tabular}
\label{tab:lookup}
\end{table}}

\subsection{Modeling Correlation by $\epsilon$-SV Regression}
\label{subsec:svr}
\RR{Consider a training set,$\{(s_1,m_1),...,$ $(s_L,m_L)\}$, where $s_i$ is the speed of the $i$-th trajectory segment and $m_i$ is the number of measurements required to recover $i$-th trajectory segment.
The $\epsilon$-SV regression determines a function $g(s)$, that given a value of $s_i$ predicts $\tilde{m_i}$, which has at most $\epsilon$ deviation from the actual $m_i$. 
The function $g$ is typically computed from 
\begin{eqnarray}
g(s)=\langle \alpha,s \rangle+b\mbox{    with $\alpha \in \mathbb{R}^n, b \in \mathbb{R}$},\label{lin_VR}
\end{eqnarray}
where $\langle.,.\rangle$ denotes the dot product within $\mathbb{R}^n$.
}

\RR{We use the matlab library LIBSVM~\cite{CC01a} to implement $\epsilon$-SVR. There are two functions: \texttt{svmtrain} and \texttt{svmpredict} for training and testing, respectively. Given the mapping of speed and number of measurements  \texttt{svmtrain}  outputs $\alpha$ and $b$. The prediction method, \texttt{svmpredict}, then use these values and some other parameters (for details please review~\cite{CC01a}) to predict the number of measurements given a new value of the mean speed.}

\RR{We used  the Radial basis function (RBF) kernel Linear Kernel which is recommended when the number of features (attributes) is relatively small. The other popular alternative is linear kernel which is however used when the number of features is sufficiently large. However, we also empirically verified that the RBF kernel performs better than the linear kernel for our datasets.
}

\subsection{Simulating Embedded Platform}
Running support vector regression on resource improvised wireless sensor nodes or mobile phones is impractical. In order to simulate the setting of these embedded devices, we have stored the mapping of speed versus compressibility in a look-up table and used linear interpolation to fetch compressibility information for a given speed. In order to find out the power consumption of various process e.g., creating projections and look-up operation on sensor node, please refer to our previous publication~\cite{Rana:2011:AAC:1966251.1966255}.

%
%
%
%

\section{Evaluation}
\label{sec:eval}

\subsection{Datasets}
We evaluate the proposed algorithm using very large datasets, containing pedestrian, and animal data. 

\subsubsection{Pedestrian Dataset}
Pedestrian traces were collected from publicly available CRAWDAD data repository~\cite{ncsu-mobilitymodels-2009-07-23}, wherein human mobility traces were collected from five different sites - 
two university campuses (NCSU and KAIST), New York City, 
Disney World (Orlando), and North Carolina state fair.
NCSU and KAIST traces were taken by $20$ and $32$ students, respectively, living in the campus dormitory. Every week, 2 or 3 randomly 
chosen students carried the GPS receivers for their daily regular 
activities. 

The New York City traces were obtained from $12$ volunteers living in 
Manhattan or its vicinity. Most of the participants have offices 
in Manhattan. Their means of 
travel include subway trains, buses and mostly walking.

The State fair track logs were collected from $8$ volunteers who 
visited a local state fair that includes many street arcades, small 
street food stands and showcases. 

The Disney World traces were obtained from $19$ volunteers 
who spent their thanksgiving or Christmas holidays in Disney World, 
Florida, USA. The GPS traces were only from 
the inside of the theme parks. The participants mainly walked 
in the parks and occasionally rode trolleys, however, we only used walking trajectories.

The data points in traces were $30$ seconds apart. Garmin GPS 60CSx handheld receivers were used for
data collection which are WAAS (Wide Area Augmentation
System) capable with a position accuracy of better than three
meters 95 percent of the time, in North America.


\subsubsection{Animal Dataset}
Animal datasets were collected from CSIRO's~\footnote{The Commonwealth Scientific and Industrial Research Organisation} sensor network deployment at Belomont, Australia. The data was collected from a virtual fencing trial involving 36 cows for 49 hours. Sampling frequency was 2 Hz.


%
%

%
\begin{figure*}
\centering
\subfigure[Pedestrian (SateFair) Dataset.]{
\includegraphics[width=0.7\linewidth]{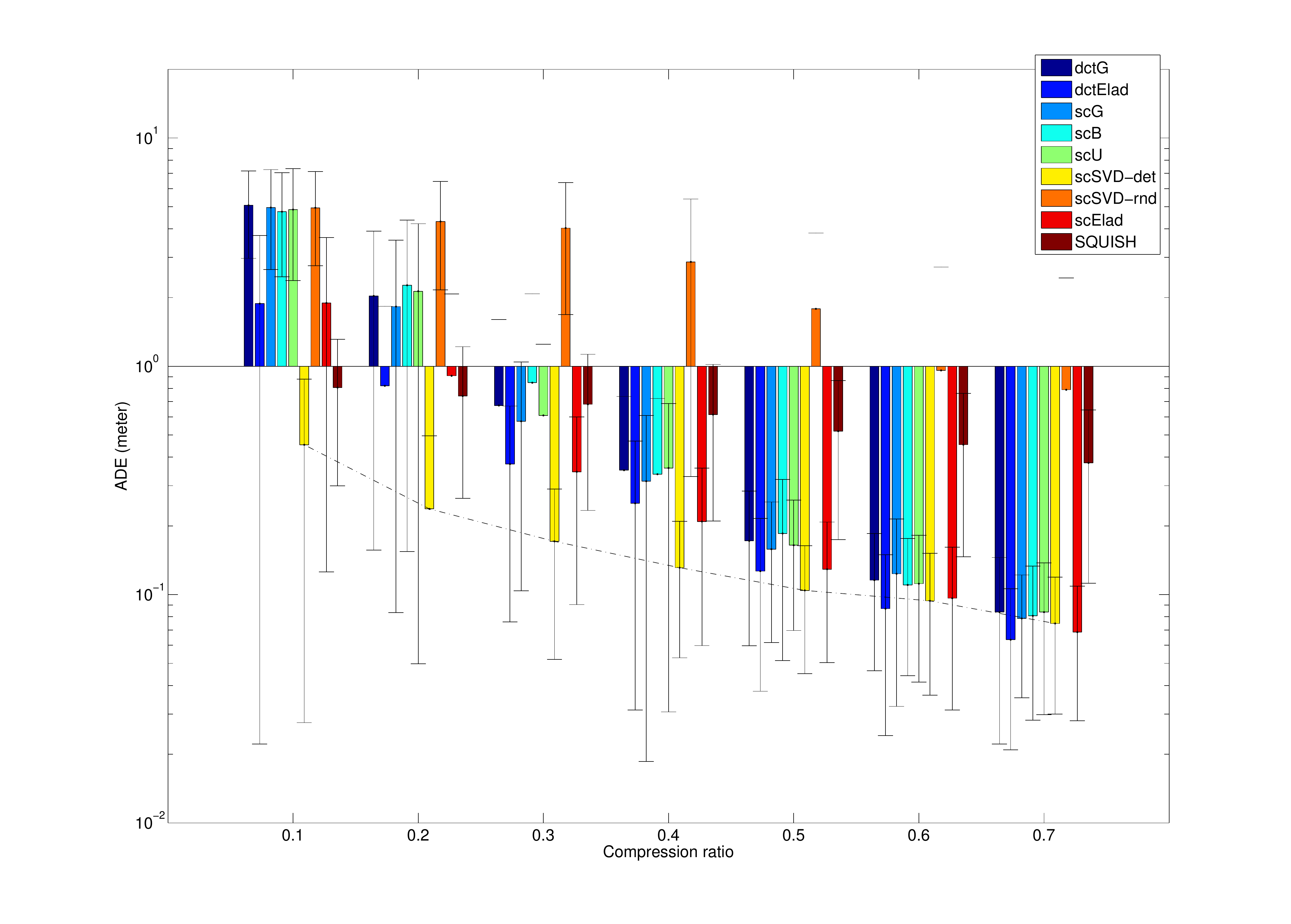}
\label{fig:NCSU}
}
\subfigure[Cattle Dataset.]{
\includegraphics[width=0.7\linewidth]{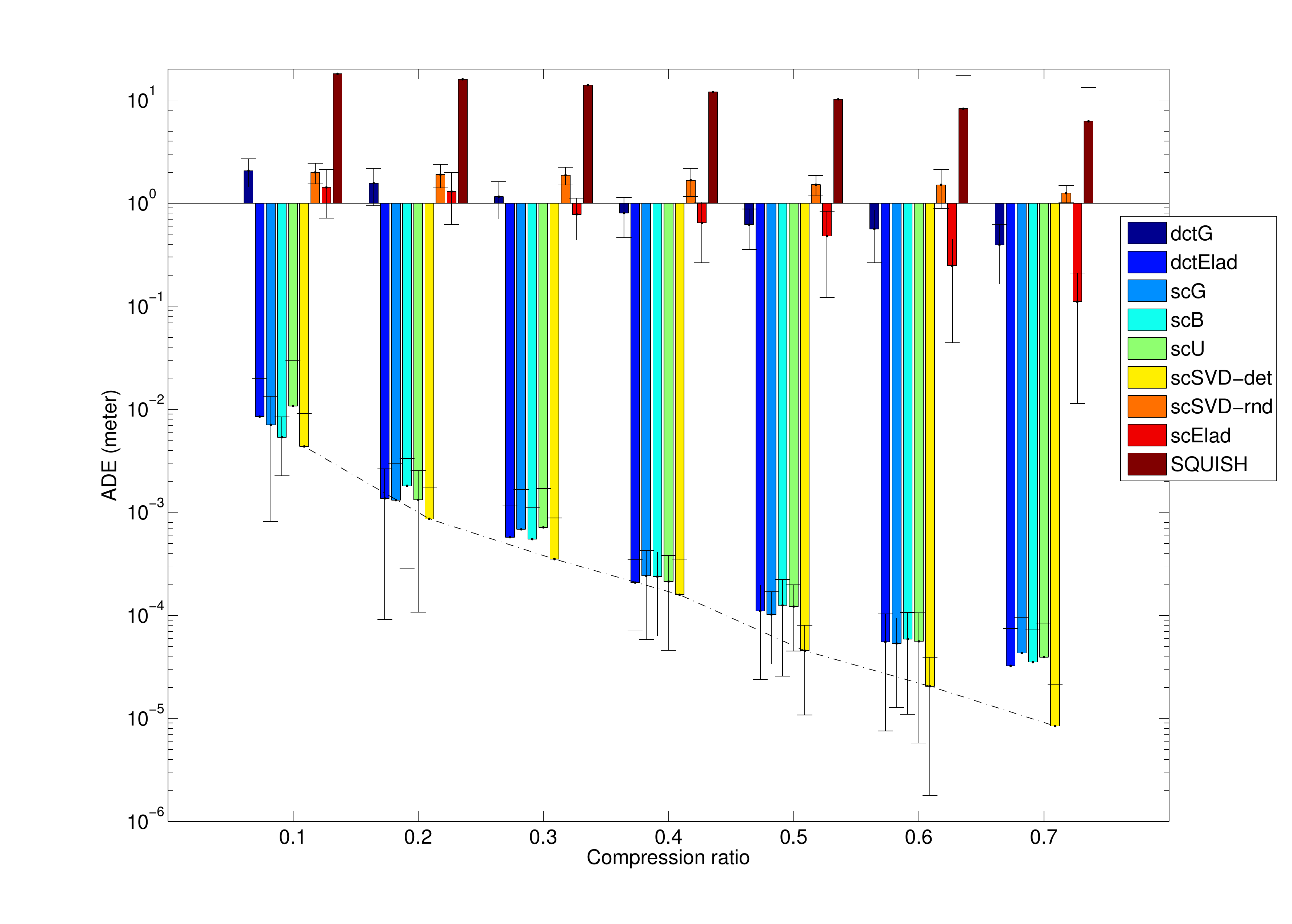}
\label{fig:cattle}
}
\caption{Comparison of compression ratio versus ADE for various methods. dctG = DCT-Gaussian pair, dctElad = DCT-Elad's optimized projection matrix pair, scG = sparse coded dictionary and Gaussian pair, scB = sparse coded dictionary and Bernoulli pair, scU = sparse coded dictionary and Unitary matrix pair, scSVD-det = sparse coded dictionary and deterministic projection matrix pair; where rows of SVD is chosen deterministically, scSVD-rnd = sparse coded dictionary and deterministic projection matrix pair; where rows of SVD is chosen randomly, scElad = sparse coded dictionary and Elad's optimized projection matrix pair.}
\label{fig:NCSUandCattle}
\end{figure*}

\subsection{Error Metric}
In this paper, we use Average Distance Error (ADE), the distance between the original and reconstructed trajectory to measure the reconstruction performance. 
Let $N^i_{\omega}, 1 \leq i \leq n$ and $E^i_{\omega}, 1 \leq i \leq n$ are the respective Northing ($y$-component) and Easting ($x$-component) data points of a trajectory segment $\omega$; and $\hat{N}^i_{\omega}, 1 \leq i \leq n$ and $\hat{E}^i_{\omega}, 1 \leq i \leq n$ are the reconstruction of $N^i_{\omega}, 1 \leq i \leq n$ and $E^i_{\omega}, 1 \leq i \leq n$, respectively.
If there are $J$ segments in a trajectory dataset, we compute ADE by, 
$\mbox{ADE=}\frac{1}{nJ}\sum_{w=1}^J\sum_{i=1}^n\sqrt{(N^i_{\omega}-\hat{N}^i_{\omega})^2+(E^i_{\omega}-\hat{E}^i_{\omega})^2}$. 


\subsection{Simulation setup}
\subsubsection{Segment Size}
We determined segment size subject to the overall speed of the object. For example, a cow does not move very rapidly, therefore we used a large segment length of 128 data points for cow. 
Pedestrian speed is intermediate to cow and vehicle, however, the sampling interval was longer: 30 seconds. The trajectory could change rapidly within the 30 seconds period, therefore, we chose a shorter segment length of 32 data points for pedestrian dataset.
\subsubsection{Datasets Filtering}
Each of the datasets required some form of filtering before we could use them. For example, in the pedestrian datasets there were GPS traces, wherein normal walking speed $6$km/hr was exceeded. We removed the corresponding  segments from the dataset.  The cow traces were continuous - not clustered by days. We divided individual's traces by days, then applied segmentations to each trace. 

\subsubsection{Dataset Processing}
We unified all the trajectories in meter units.
Northing and Easting in meters were 7 and 6 digit numbers with two decimal points, respectively. We observed that $\ell_1$-minimization is not particularly good with large numbers. For each segment we subtracted the mean of the segment from all data points of the segment to make up the numbers with smaller digits.

\subsubsection{Training and Tetsting sets}
The pedestrian datasets were collected from $5$ different sites. We learned separate dictionary for each sites. Note that although there should be a commonality among these 5 different sites data due to regular walking speed and patterns, due to the structure of the sites (road structure etc.), the trajectory patterns were very different. For cow datasets, we learned separate dictionaries for individual cows, due to varying mobility pattens. We used $80\%$ of the segments for training i.e., for learning the dictionary, and used remaining $20\%$ for testing the performance of the compression algorithm. 

\subsubsection{Benchmark Candidates}

We contrast the performance of the sparse coded dictionary with Discrete Cosine Transform (DCT), since in our previous study~\cite{conf/ewsn/RanaHWC11} we observed that DCT is the most suitable sparsifier for trajectory dataset. We compare the performance of the proposed projection matrix with four different random projection matrices suggested by the compressive sensing literature: 1. Gaussian, 2. Bernoulli, 3. Unitary, and 4. incoherent matrix (see Section~\ref{sec:saprsityRip}).  Note that Gaussian matrix is also an incoherent matrix, therefore, we do not present incoherent matrix as a separate entity. In the plot legend ``sc'' refers to Sparse Coded, ``G'' refers to Gaussian, ``B'' refers to Bernoulli and ``U'' refers to Unitary and ``SVD'' refers to the single value decomposition. ``scSVD-Det'' is our proposed deterministic construction of projection matrix and sparse coded dictionary pair.  ``scSVD-Rnd'' refers to the random construction of projection matrix proposed in~\cite{zhou2009non}. In order to avoid cluttering the image, we only use Gaussian matrix paired with DCT (``dctG''), the other two matrices (B and U) produce similar results. 

We also contrast the proposed ``scSVD-Det'' with the method proposed by Michael Elad in~\cite{elad2007optimized}. This work is very closely related to our work wherein the author optimizes the projection matrix and demonstrates that the optimized matrix performs significantly better than the random projection matrices (More details are provided in the related work section). In order to perform a fair comparison, we use Elad's projection matrix with our sparse coded dictionary (``scElad''). However, we also use the projection matrix with predefined DCT basis (``dctElad'').

Last but not the least we compare the compression performance of the proposed framework with SQUISH~\cite{muckell2011squish}, which is a powerful GPS compression algorithm recently proposed by Muckell et al.~\cite{muckell2011squish}. SQUISH works on the principles of Synchronous Euclidian Distance and reported to perform better than the prominent methods e.g., Uniform Sampling, Online Dead Reckoning, and Online Douglas-Peucker, when compression ratio is small.

Note that other powerful algorithms e.g., LZW, LZ77 are mainly suitable for text compression, therefore, we could not compare with these methods.

\subsubsection{Simulation Environment}
The simulations were written in Matlab 2010b. 

\subsubsection{Compression Prediction}
\RR{Recall that we use support vector regression to predict the number of measurements from the mean speed of a trajectory segment.  In this section, we report two key results related to the prediction: First in Fig.~\ref{fig:meanpredictionerrro}, we show the prediction error. We use training datasets to train SVR and use the test datasets to test the prediction accuracy. We compute mean prediction error using the following formula:
\begin{eqnarray}
\mbox{\bf {Percentage prediction error} } \nonumber\\=\frac{1}{\mbox{total segments}} \sum\frac{\mbox{(real - predicted) projecitons}}{\mbox{segment length}}\%
\label{eqn:predictionAccuracy}
\end{eqnarray}}

\RR{The other quantity we report is percentage transmission savings due to adaptive compression. We contrast the adaptive approach with a conservative approach that considers the historical maximum value of number of projections. We compute the percentage transmission savings using the following formula. 
\begin{eqnarray}
\mbox{{\bf Percentage transmission savings} } \nonumber\\=\frac{1}{\mbox{total segments}} \sum{\frac{\mbox{(max - predicted) projecitons}}{\mbox{segment length}}}\%
\label{eqn:spaceSavings}
\end{eqnarray}}

%
%

\subsection{Simulation Results}
\label{sec:simulationResult}
The results of compression ratio versus reconstruction error are plotted in Fig.~\ref{fig:NCSUandCattle}. We report the results for one pedestrian site and one cattle, since they represent the population quite well. However, since we have pedestrian datasets from different  sites, we report the results from rest four pedestrian sites in the Appendix A. In order to be clearly visible we have drawn a line with the bar representing the proposed method (scSVD-det) in this paper.

\subsubsection{Comparison amongst Projection Matrices and Dictionaries}   

We start with the comparison of custom versus predefined dictionaries.
We choose dctG and scG for this comparison. Across all the pedestrian and cattle datasets, there is a very insignificant performance difference between these two pairs. 
%

To contrast deterministic with random projection matrices we choose scG and scSVD. For pedestrian datasets, upto compression ratio $0.3$, scSVD performs significantly better compared to scG: the reconstruction error is at least $10$ times small. However, when compression ratio rises above $0.3$, the performance difference between these two pairs gets narrower.
Note that performance of scB and scU are quite similar to that of scG, therefore, comparison with scG is sufficient.  For the cattle datasets, when the compression ratio is less than equal to $0.3$, reconstruction error given by scSVD-det is almost 5 times smaller than scG; however, similar to the pedestrian dataset, when the compression ratio goes above $0.3$, the gap diminishes gradually. 

%

\begin{figure*}
\centering
\subfigure[]{
\includegraphics[width = 0.5\linewidth]{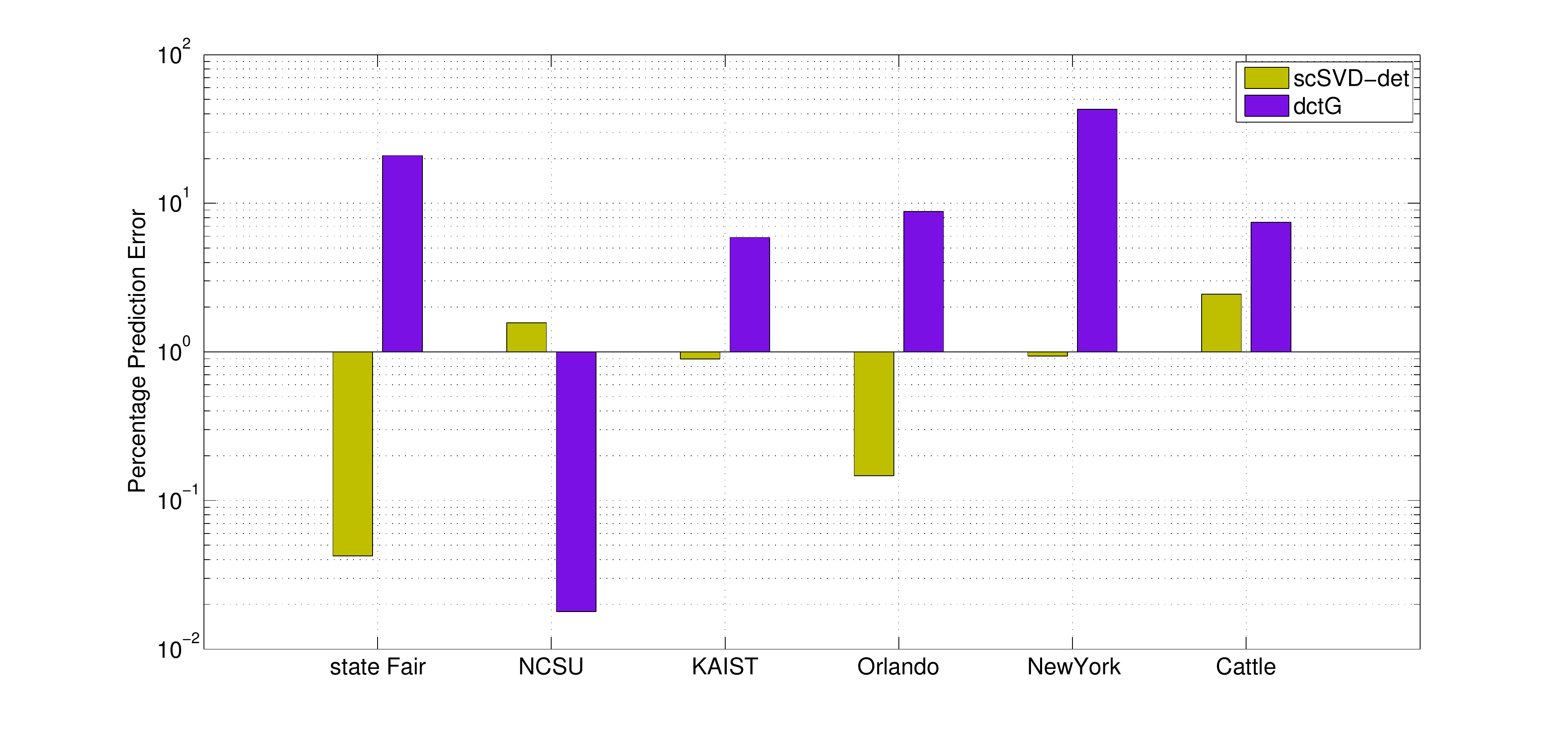}
\label{fig:meanpredictionerrro}
}
\subfigure[]{
\includegraphics[width = 0.45\linewidth]{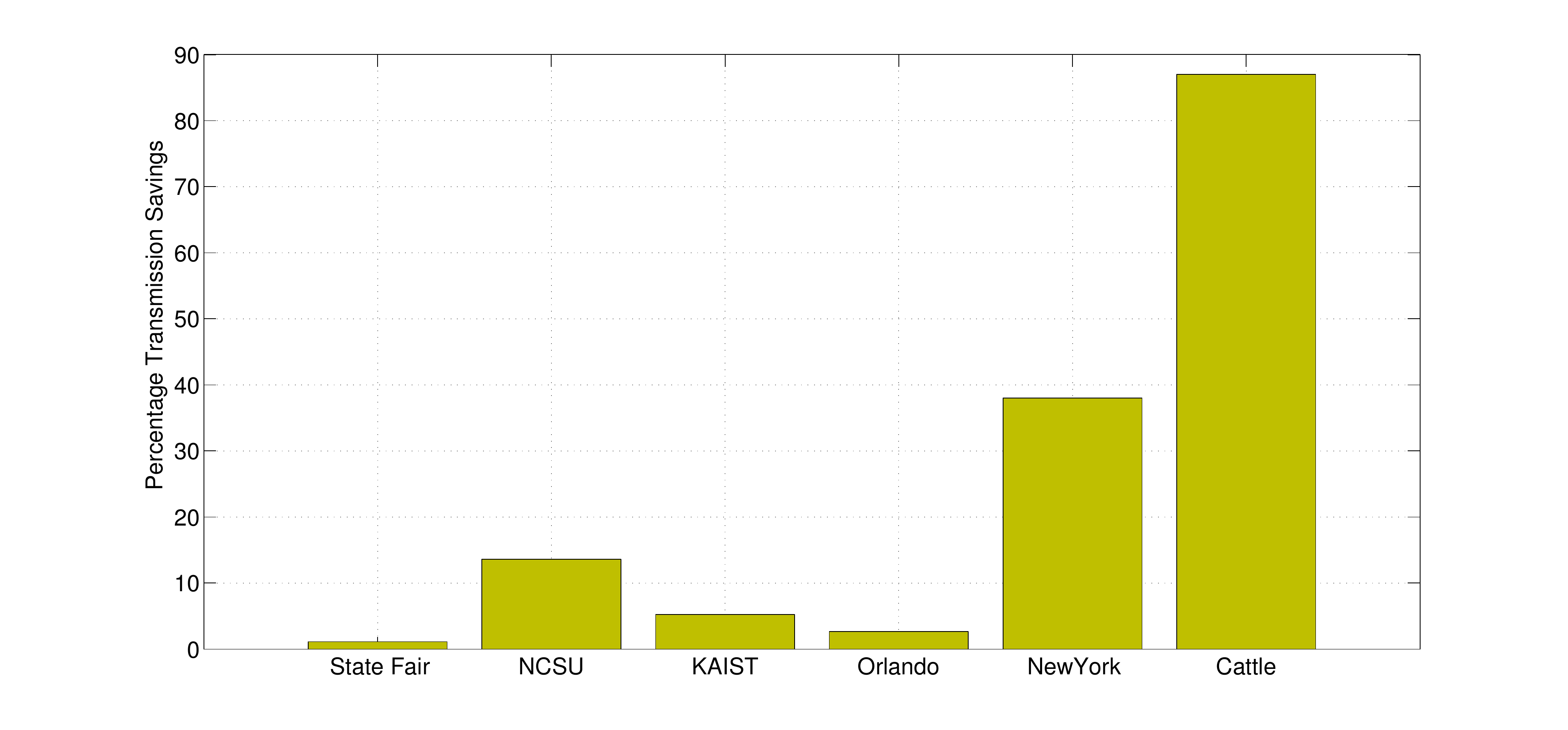}
\label{fig:spaceSavings}
}
\caption{Performance of adaptive compression.}
\end{figure*}


From Fig.~\ref{fig:NCSUandCattle} it is conclusive that the scSVD-det performs significantly better than scSVD-rnd. In particular, scSVD-rnd performs worse than all other methods we use in the paper.
The comparison of scSVD-det with dctElad and scElad reveals that for smaller compression ratio specifically, compression ratio up to  $0.3$, scSVD-det performs significantly better than dctElad and scElad. For example, when compression ratio is $0.1$, for pedestrian dataset, scSVD-det produces 10 cm error, whereas scElad or dctElad produces minimum 1 meter error (maximum $10$m error). However, for cattle dataset, the scElad or dctELad do not perform as good as the pedestrian dataset, therefore, the difference between scSVD-det and scElad/dctElad is even higher. For example, for cattle datatset, when the compression ratio is 0.1, scSVD-det has 1 centimeter error, whereas scElad has 3 meter error, however, dctElad has few centimeter error. Note that scSVD-det performs consistently better with the sparse coded dictionary for both pedestrian and cattle dataset. 

SQUISH performs well for the pedestrian dataset, but not for the cattle dataset. For pedestrian dataset, except for the scSVD-det, SQUISH performs better than most  other methods, especially when the compression ratio is small. For cattle dataset, performance of SQUISH is bad compared to any other methods.

\RR{\subsubsection{Conservative versus Adaptive Compression}
In Fig.~\ref{fig:meanpredictionerrro} we demonstrate the prediction performance of our proposed method. We compute the percentage prediction error using \eqref{eqn:predictionAccuracy}. The maximum error is below 3\% (cattle), which confirms our highly accurate prediction performance. }

\RR{We also contrast the performance of the proposed method with our previously proposed compression framework, wherein we used predefined dictionary and projection matrix pair~\cite{conf/ewsn/RanaHWC11}. In particular we used dctG pair for compression and support vector regression for prediction. Therefore, in Fig.~\ref{fig:meanpredictionerrro}, we contrast ``scSVD-Det''  with ``dctG''. We observe that our newly proposed compression framework significantly outperforms  the previously proposed one. For pedestrian dataset, the maximum improvement is more than 20\% (State Fair) and the minimum improvement is more than 5\% (KAIST). The representative of cattle dataset offers more than 5\% improvement.}

\RR{We compute percentage transmission savings using~\eqref{eqn:spaceSavings}. Adapting the number of projections to the mean speed of the object, we can achieve significant transmission savings as depicted in Fig.~\ref{fig:spaceSavings}. Maximum transmission savings for pedestrian dataset is about 40\% for NewYork city data and minimum savings is for State Fair - 2\%. Transmission savings for cattle dataset 
is quite high - about 85\%. }

\subsubsection{Results Summary}
We can summarize the results as follows:
\begin{enumerate}
\item The performance of the proposed deterministic projection matrix is significantly better compared to the predefined and other proposed                                                                                                                                                                                                                                                  alternatives in the literature. For example, scSVD-det produces 10 times less error compared to dctG and scElad. 

\item The compression performance gets even better when the proposed projection matrix and dictionary is applied jointly with the proposed adaptive compression framework.  For example, for cattle dataset the adaptive compression using scSVD-det offers 85\% improvement over dctG, whereas without the adaptive part the improvement is only 10 times.
                                                                                                                                                                                                                                                                                                           
\item The proposed compression algorithm is better than the existing trajectory compression algorithm - SQUISH, which has been reported to perform better than the prominent trajectory methods e.g., Uniform Sampling, Online Dead Reckoning, and Online Douglas-Peucker when compression ratio is small.

\end{enumerate}

\section{Related Work}
\label{sec:related}

The solution provided in this paper spans two key aspects: 1. adapting the compression 2. deterministic construction of projection matrix. Therefore, we will evaluate the literature mainly in these two aspects. However, we will briefly touch upon the trajectory compression algorithms proposed in wireless sensor networks.

\subsection{Adaptive Compression in Wireless Sensor Networks}
The adaptive compression algorithms proposed so far in the wireless sensor network mainly adapt compression for energy savings. Most of the algorithms proposed in the past mainly consider slowly changing natural phenomena, which intrinsically require relatively low sampling. Therefore, bandwidth conservation has got secondary focus compared to energy conservation.  For example, in~\cite{1208942} authors propose an adaptive compassion algorithm, wherein compression is adapted at the sensing node by analyzing the correlation in a centralized  data store. Since the approach require central server to node communication, it is suitable for slowly changing phenomena e.g., soil moisture. However, we consider trajectory with as high as 2 Hz sampling rate, therefore, such technique may result in enormous node to base communication causing quick depletion of the sensor node battery. 

In~\cite{ali2011adaptive}  authors propose a 3-stage adaptation framework wherein in  stage 1, correlated sensor nodes form small 1-hop clusters based on a short history of the attribute values to exploit strong local correlation. In Stage 2,  the temporal correlations is exploited by constructing the models on a small number of clusters, referred to as master clusters. Each constructed model is limited to the respective master clusters and approximates the sampled values of all the member sensor nodes of the master cluster. In Stage 3, the master cluster sends the model to its neighboring clusters. The cluster members fit the received model to their sampled values and accordingly either accept the model or reject the model. The clusters accepting the model merge to form a correlated region or larger clusters and further propagate the model to their neighboring clusters. Following this scheme, only a small set of the models constructed on master clusters can approximate the entire network both in space and time.  However, similar to the work in~\cite{1208942}, this method is well suited to slowly changing phenomena; when used for trajectory compression, this algorithm will require enormous internode communication, resulting in quick decay of the node energy.

Some other adaptive compression algorithms, although do not require lots of inter-node communication, however, require large number of on-node processing.  For example, in~\cite{dong2006adaptive}  authors propose an adaptive wavelet compression algorithm for wireless sensor networks. In the proposed method each receiving sensor computes the compression ratio, and calculates the total energy dissipation (using both computation and communication energy models) to make a decision about whether to increases wavelet transform level or to keep the present level. Then, the sensor, runs wavelet compression with next transform level to compute the new compression ratio, computes new value of total energy dissipation and compares it with the old value. The above steps will be repeated if the new energy estimate is smaller than old estimate and wavelet transform level is less than some maximum allowed value. After this operation, the nodes  transmit data to the central nodes applying the computed wavelet transform level. This method will involve enormous computation given that for each trajectory segment it has to iterate multiple times to determine the optimal transform level for the best compression and energy trade-off.

Similar problem will be experienced in the algorithm proposed in~\cite{puthenpurayil2007energy}, which  employs a feedback approach in which the compression ratio is compared to a pre-determined threshold. The compression model used in the previous frame can be retained and used for the next frame, if compression ratio is greater than the predefined threshold. Otherwise, the adaptive operation of the system will produce a new compression model. 

A slightly different adaptive compression principle is applied in the algorithm proposed in \cite{akyildiz2006state}.  Authors design an on-line adaptive algorithm that dynamically makes compression decisions to accommodate the changing state of WSNs. In the algorithm, a queueing model is adopted to estimate the queueing behavior of sensors with the assistance of only local information of each sensor node. By using the queueing model, the algorithm predicts the compression effect on the average packet delay and performs compression only when it can reduce the packet delay. This algorithm is quite elegant since it does not require lots of on node processing and intra-node communications, however, this algorithm may not be suitable for any trajectory in general. Instead, this algorithm will be suitable for those trajectories, where objects keep stationary for substantial amount of time, therefore, compression will be applied only when they are moving. Note that our proposed method is more general. The compression ratio is adapted to the speed of the object, therefore, when the object is not moving maximum compression will be achieved, and as the object starts moving instead of maintaining a common compression ratio, we adapt compression ration to the speed. 

Finally, in~\cite{kolo2012adaptive} authors present an adaptive lossless data compression (ALDC) algorithm for wireless sensor networks. The data sequence to be compressed is partitioned into blocks, and the optimal compression scheme is applied for each block. However, the proposed algorithm is lossless, therefore, it is not robust to data loss of the wireless sensor network platform.

\subsection{Deterministic Construction of Projection Matrix}
Elad in~\cite{elad2007optimized} and Julio et al. in~\cite{duarte2009learning} have optimized projection matrix to achieve better compression ratio. Elad has defined a new mutual coherence, which describes the correlation between the dictionary and projection matrix. The smaller the mutual coherence, the better the compression performance. Elad has minimized the mutual coherence with respect to the projection matrix - keeping the dictionary fixed. In addition to just optimizing the projection matrix, Julio et al. has optimized the dictionary simultaneously.  
In particular, Julio uses recently proposed K-SVD algorithm proposed in \cite{aharon2005k} to learn dictionary and then jointly optimize the dictionary and projection matrix by maximizing the number of orthogonal columns in their product. We use SPAMS to learn the dictionary, which is different to K-SVD. In addition, in order to optimize the projection matrix we obtain a special singular value decomposition of the dictionary, which naturally produces low coherence projection matrix and dictionary pair. We contrasted our work with that of Elad's, however, we had difficulties to run Julio's method for our trajectory dataset.

\subsection{Trajectory/GPS Compression Algorithms for Wireless Sensor Networks}
A very small number of work can be found in the literature, which propose  trajectory compression algorithm for wireless sensor network or other embedded platform. One of such algorithm is the compression algorithm proposed in~\cite{dttc}. This algorithm performs recursive 
segmentation of the trajectory, 
until a trajectory segment can be modelled with an interpolation function with a small error. 
Compression is achieved by only transmitting 
the relevant parameters of the interpolation function. 
represents trajectory segments by some linear or polynomial functions and achieve compression by transmitting  the compressed functions instead of trajectory data points. 
However, due to large computation requirement the proposed compression algorithm is not suitable for real-time compression.

In~\cite{ghica2010trajectory} authors propose a trajectory compression algorithm which uses various line simplification methods, for example, Dead-Reckoning and the Douglas-Peuker algorithm, and a variant of a CG-based optimal algorithm for polyline reduction. In particular, the authors also propose a hybrid approachm, which combines some of the above methods. Note that out of the three methods, Douglas-Peuker is most popular. In our previous work, we have already shown that the non-optimized version of our projection matrix already performs better than the improved Douglas-Peuker method proposed by Meratina et al~\cite{meratnia2004spatiotemporal}.



\section{Conclusion and Future Work}
\label{sec:conclusion}
We have proposed an adaptive compression framework to achieve improved compression performance underpinning the theory of compressive sensing and support vector regression. We adapt the compression subject to the mean speed of the object to improve the compression gain. We learn a sparsifying dictionary from the dataset using the theory of sparse coding and construct the projection matrix from the dictionary by applying singular value decomposition on the dictionary.   We case study GPS trajectory spanning pedestrian and animal trajectories across three different countries involving more than 120 subjects and conclude that, 
\begin{enumerate}
\item adaptive compression is very useful to increase the trajectory compression performance,
\item deterministic construction of projection matrix  is more suitable compared to the predefined random matrices to achieve better trajectory compression performance.
\end{enumerate}
%
%



\RR{Due to the large sampling interval, in this paper we only considered low speed trajectories. In our future study, we aim to obtain vehicle datasets with possibly smaller sampling intervals and validate the performance of compressive sensing for compressing high speed object trajectory.}

\bibliographystyle{plain}
\bibliography{reference_3,sigproc}

\begin{thebibliography}{10}

\bibitem{miniLZo}
\url{http://www.oberhumer.com/opensource/lzo/}, 2011.

\bibitem{aharon2005k}
Michal Aharon, Michael Elad, and Alfred Bruckstein.
\newblock K-svd: Design of dictionaries for sparse representation.
\newblock {\em Proceedings of SPARS}, 5:9--12, 2005.

\bibitem{akyildiz2006state}
Ian~F Akyildiz, Dario Pompili, and Tommaso Melodia.
\newblock State-of-the-art in protocol research for underwater acoustic sensor
  networks.
\newblock In {\em Proceedings of the 1st ACM international workshop on
  Underwater networks}, pages 7--16. ACM, 2006.

\bibitem{ali2011adaptive}
Azad Ali, Abdelmajid Khelil, Piotr Szczytowski, and Neeraj Suri.
\newblock An adaptive and composite spatio-temporal data compression approach
  for wireless sensor networks.
\newblock In {\em Proceedings of the 14th ACM international conference on
  Modeling, analysis and simulation of wireless and mobile systems}, pages
  67--76. ACM, 2011.

\bibitem{BaraniukDavenportDeVoreWakin:08}
R.~Baraniuk, M.~Davenport, R.~DeVore, and M.~Wakin.
\newblock A simple proof of the restricted isometry property for random
  matrices.
\newblock {\em Constr Approx}, 28(3):253--263, 2008.

\bibitem{Bertsekas99}
Dimitri~P. Bertsekas and Dimitri~P. Bertsekas.
\newblock {\em {Nonlinear Programming}}.
\newblock Athena Scientific, 2nd edition, September 1999.

\bibitem{bourgain1985lipschitz}
Jean Bourgain.
\newblock On lipschitz embedding of finite metric spaces in hilbert space.
\newblock {\em Israel Journal of Mathematics}, 52(1-2):46--52, 1985.

\bibitem{CaiWangXu:10b}
T.T. Cai, Lie Wang, and Guangwu Xu.
\newblock New bounds for restricted isometry constants.
\newblock {\em Information Theory, IEEE Transactions on}, 56(9):4388 --4394,
  sept. 2010.

\bibitem{Candes2008:Introduction}
E.~J. Candes and M.~B. Wakin.
\newblock An introduction to compressive sampling.
\newblock {\em IEEE Signal Processing Magazine}, 25(2):21--30, 2008.

\bibitem{CandesTao:06}
E.J. Candes and T.~Tao.
\newblock Near-optimal signal recovery from random projections: Universal
  encoding strategies?
\newblock {\em Information Theory, IEEE Transactions on}, 52(12):5406 --5425,
  dec. 2006.

\bibitem{Candes:08}
Emmanuel~J. Candes.
\newblock The restricted isometry property and its implications for compressed
  sensing.
\newblock {\em Comptes Rendus Mathematique}, 346(9–10):589 -- 592, 2008.

\bibitem{CandesEldarNeedellRandall:11}
Emmanuel~J. Candes, Yonina~C. Eldar, Deanna Needell, and Paig Randall.
\newblock Compressed sensing with coherent and redundant dictionaries.
\newblock {\em Applied and Computational Harmonic Analysis}, 31(1):59--73,
  2011.

\bibitem{CC01a}
Chih-Chung Chang and Chih-Jen Lin.
\newblock {LIBSVM}: A library for support vector machines.
\newblock {\em ACM Transactions on Intelligent Systems and Technology},
  2:27:1--27:27, 2011.
\newblock Software available at \url{http://www.csie.ntu.edu.tw/~cjlin/libsvm}.

\bibitem{1208942}
J.~Chou, D.~Petrovic, and Kannan Ramachandran.
\newblock A distributed and adaptive signal processing approach to reducing
  energy consumption in sensor networks.
\newblock In {\em INFOCOM 2003. Twenty-Second Annual Joint Conference of the
  IEEE Computer and Communications. IEEE Societies}, volume~2, pages 1054--1062
  vol.2, 2003.

\bibitem{DaviesGribonval:09}
M.E. Davies and R.~Gribonval.
\newblock Restricted isometry constants where $\ell ^{p}$ sparse recovery can
  fail for $0 \ll p \leq 1$.
\newblock {\em Information Theory, IEEE Transactions on}, 55(5):2203 --2214,
  may 2009.

\bibitem{dong2006adaptive}
Hui Dong, Jiangang Lu, and Youxian Sun.
\newblock Adaptive distributed compression algorithm for wireless sensor
  networks.
\newblock In {\em Innovative Computing, Information and Control, 2006.
  ICICIC'06. First International Conference on}, volume~3, pages 283--286.
  IEEE, 2006.

\bibitem{duarte2009learning}
Julio~M Duarte-Carvajalino and Guillermo Sapiro.
\newblock Learning to sense sparse signals: Simultaneous sensing matrix and
  sparsifying dictionary optimization.
\newblock {\em Image Processing, IEEE Transactions on}, 18(7):1395--1408, 2009.

\bibitem{elad2007optimized}
Michael Elad.
\newblock Optimized projections for compressed sensing.
\newblock {\em Signal Processing, IEEE Transactions on}, 55(12):5695--5702,
  2007.

\bibitem{ghica2010trajectory}
Oliviu Ghica, Goce Trajcevski, Ouri Wolfson, Ugo Buy, Peter Scheuermann, Fan
  Zhou, and Dennis Vaccaro.
\newblock Trajectory data reduction in wireless sensor networks.
\newblock {\em INTERNATIONAL JOURNAL OF NEXT-GENERATION COMPUTING}, 1(1), 2010.

\bibitem{goyal}
V.~K. Goyal, A.~K. Fletcher, and S.~Rangan.
\newblock Compressive sampling and lossy compression.
\newblock {\em Signal Processing Magazine, IEEE}, 25(2):48--56, March 2008.

\bibitem{kolo2012adaptive}
Jonathan~Gana Kolo, S~Anandan Shanmugam, David Wee~Gin Lim, Li-Minn Ang, and
  Kah~Phooi Seng.
\newblock An adaptive lossless data compression scheme for wireless sensor
  networks.
\newblock {\em Journal of Sensors}, 12, 2012.

\bibitem{lee07}
Honglak Lee, Alexis Battle, Rajat Raina, and Andrew~Y. Ng.
\newblock {Efficient sparse coding algorithms}.
\newblock In {\em In NIPS}, pages 801--808, 2007.

\bibitem{Mairal:2009:ODL:1553374.1553463}
Julien Mairal, Francis Bach, Jean Ponce, and Guillermo Sapiro.
\newblock Online dictionary learning for sparse coding.
\newblock In {\em Proceedings of the 26th Annual International Conference on
  Machine Learning}, ICML '09, pages 689--696, New York, NY, USA, 2009. ACM.

\bibitem{marcelloni2009efficient}
Francesco Marcelloni and Massimo Vecchio.
\newblock An efficient lossless compression algorithm for tiny nodes of
  monitoring wireless sensor networks.
\newblock {\em The Computer Journal}, 52(8):969--987, 2009.

\bibitem{Marcelloni20101924}
Francesco Marcelloni and Massimo Vecchio.
\newblock Enabling energy-efficient and lossy-aware data compression in
  wireless sensor networks by multi-objective evolutionary optimization.
\newblock {\em Information Sciences}, 180(10):1924 -- 1941, 2010.
\newblock <ce:title>Special Issue on Intelligent Distributed Information
  Systems</ce:title>.

\bibitem{MendelsonPajorTomczak-Jaegermann:08}
S.~Mendelson, A.~Pajor, and N.~Tomczak-Jaegermann.
\newblock Uniform uncertainty principle for bernoulli and subgaussian
  ensembles.
\newblock {\em Constr Approx}, 28(3):277--289, 2008.

\bibitem{meratnia2004spatiotemporal}
Nirvana Meratnia and A~Rolf.
\newblock Spatiotemporal compression techniques for moving point objects.
\newblock In {\em Advances in Database Technology-EDBT 2004}, pages 765--782.
  Springer, 2004.

\bibitem{MoLi:11}
Qun Mo and Song Li.
\newblock New bounds on the restricted isometry constant $\delta_{2k}$.
\newblock {\em Applied and Computational Harmonic Analysis}, 31(3):460--468,
  2011.

\bibitem{muckell2011squish}
Jonathan Muckell, Jeong-Hyon Hwang, Vikram Patil, Catherine~T Lawson, Fan Ping,
  and SS~Ravi.
\newblock Squish: an online approach for gps trajectory compression.
\newblock In {\em Proceedings of the 2nd International Conference on Computing
  for Geospatial Research \& Applications}, page~13. ACM, 2011.

\bibitem{Osborne99anew}
M.~R. Osborne, Brett Presnell, and B.A. Turlach.
\newblock A new approach to variable selection in least squares problems, 1999.

\bibitem{puthenpurayil2007energy}
Sebastian Puthenpurayil, Ruirui Gu, and Shuvra~S Bhattacharyya.
\newblock Energy-aware data compression for wireless sensor networks.
\newblock In {\em Acoustics, Speech and Signal Processing, 2007. ICASSP 2007.
  IEEE International Conference on}, volume~2, pages II--45. IEEE, 2007.

\bibitem{Rana:2011:AAC:1966251.1966255}
Rajib Rana, Wen Hu, Tim Wark, and Chun~Tung Chou.
\newblock An adaptive algorithm for compressive approximation of trajectory
  (aacat) for delay tolerant networks.
\newblock In {\em Proceedings of the 8th European conference on Wireless sensor
  networks}, EWSN'11, pages 33--48, Berlin, Heidelberg, 2011. Springer-Verlag.

\bibitem{conf/ewsn/RanaHWC11}
Rajib~Kumar Rana, Wen Hu, Tim Wark, and Chun~Tung Chou.
\newblock An adaptive algorithm for compressive approximation of trajectory
  (aacat) for delay tolerant networks.
\newblock In Pedro~Jos� Marr�n and Kamin Whitehouse, editors, {\em EWSN},
  volume 6567 of {\em Lecture Notes in Computer Science}, pages 33--48.
  Springer, 2011.

\bibitem{ncsu-mobilitymodels-2009-07-23}
Injong Rhee, Minsu Shin, Seongik Hong, Kyunghan Lee, Seongjoon Kim, and Song
  Chong.
\newblock {CRAWDAD} data set ncsu/mobilitymodels (v. 2009-07-23).
\newblock Downloaded from http://crawdad.cs.dartmouth.edu/ncsu/mobilitymodels,
  July 2009.

\bibitem{RudelsonVershynin:08}
Mark Rudelson and Roman Vershynin.
\newblock On sparse reconstruction from fourier and gaussian measurements.
\newblock {\em Communications on Pure and Applied Mathematics},
  61(8):1025--1045, 2008.

\bibitem{Sadler2006}
Christopher~M. Sadler and Margaret Martonosi.
\newblock Data compression algorithms for energy-constrained devices in delay
  tolerant networks.
\newblock In {\em SenSys}, pages 265--278. ACM, 2006.

\bibitem{schoellhammer2004lightweight}
Tom Schoellhammer, Ben Greenstein, Eric Osterweil, Michael Wimbrow, and Deborah
  Estrin.
\newblock Lightweight temporal compression of microclimate datasets.
\newblock 2004.

\bibitem{Wark:2007:TAT:1262537.1262569}
Tim Wark, Peter Corke, Pavan Sikka, Lasse Klingbeil, Ying Guo, Chris Crossman,
  Phil Valencia, Dave Swain, and Greg Bishop-Hurley.
\newblock Transforming agriculture through pervasive wireless sensor networks.
\newblock {\em IEEE Pervasive Computing}, 6(2):50--57, April 2007.

\bibitem{dttc}
Yingqi Xu and Wang-Chien Lee.
\newblock Dttc: Delay-tolerant trajectory compression for object tracking
  sensor networks.
\newblock In {\em Proceedings of the IEE International Conference on Sensor
  Networks, Ubiquitous, and Trustworthy Computing (SUTC,06)}, pages 436--445,
  2006.

\bibitem{Carin:12}
Mingyuan Zhou, Haojun Chen, J.~Paisley, Lu~Ren, Lingbo Li, Zhengming Xing,
  D.~Dunson, G.~Sapiro, and L.~Carin.
\newblock Nonparametric bayesian dictionary learning for analysis of noisy and
  incomplete images.
\newblock {\em Image Processing, IEEE Transactions on}, 21(1):130 --144, jan.
  2012.

\bibitem{zhou2009non}
Mingyuan Zhou, Haojun Chen, John Paisley, Lu~Ren, Guillermo Sapiro, and
  Lawrence Carin.
\newblock Non-parametric bayesian dictionary learning for sparse image
  representations.
\newblock 2009.

\end{thebibliography}

\appendices
\section{}
\label{appen:A}

\label{appen:A}
\begin{figure*}
\centering
\includegraphics[width=0.7\linewidth]{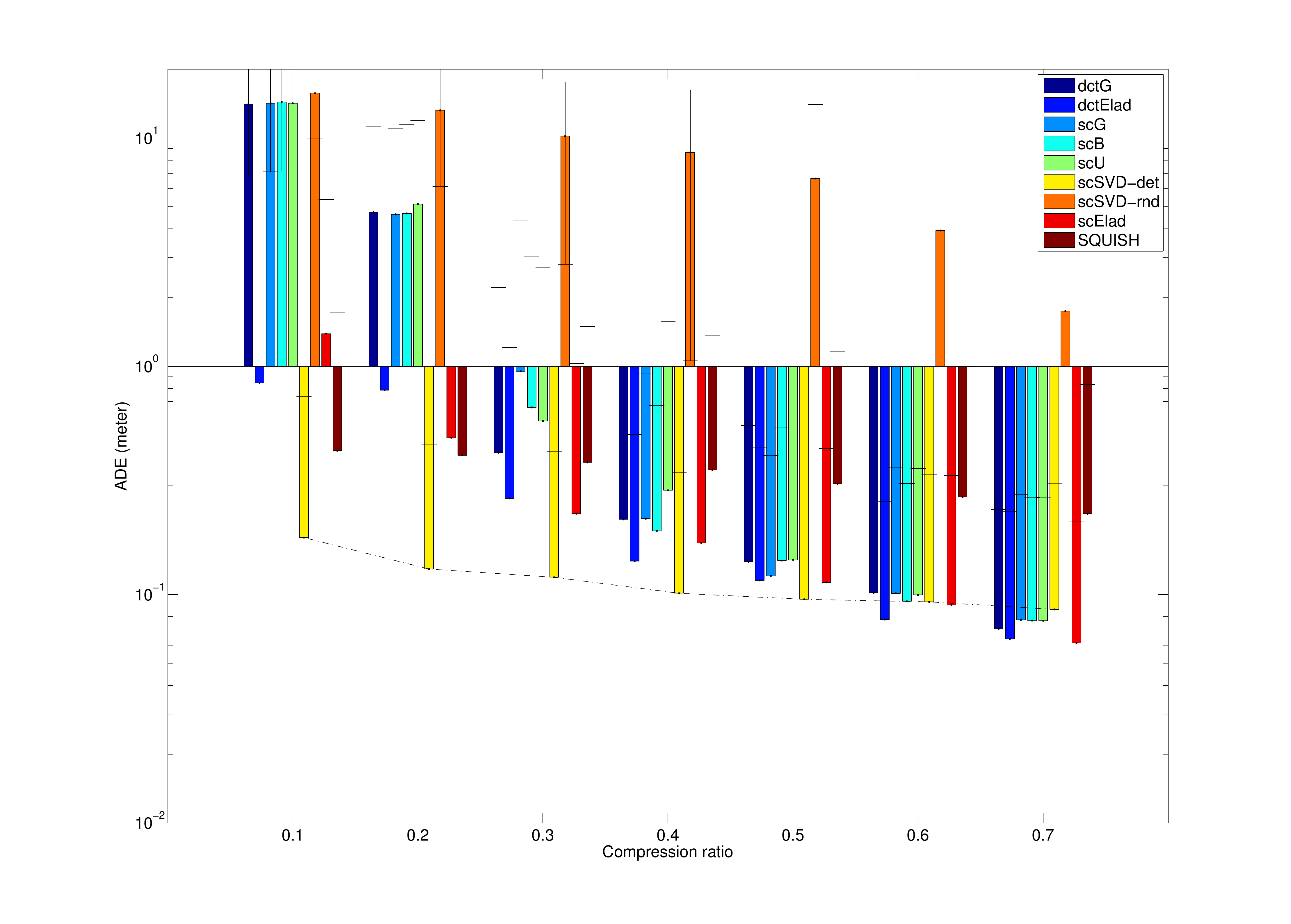}
\caption{NCSU}
\label{fig:NCSU}
\end{figure*}

\begin{figure*}
\centering
\includegraphics[width=0.7\linewidth]{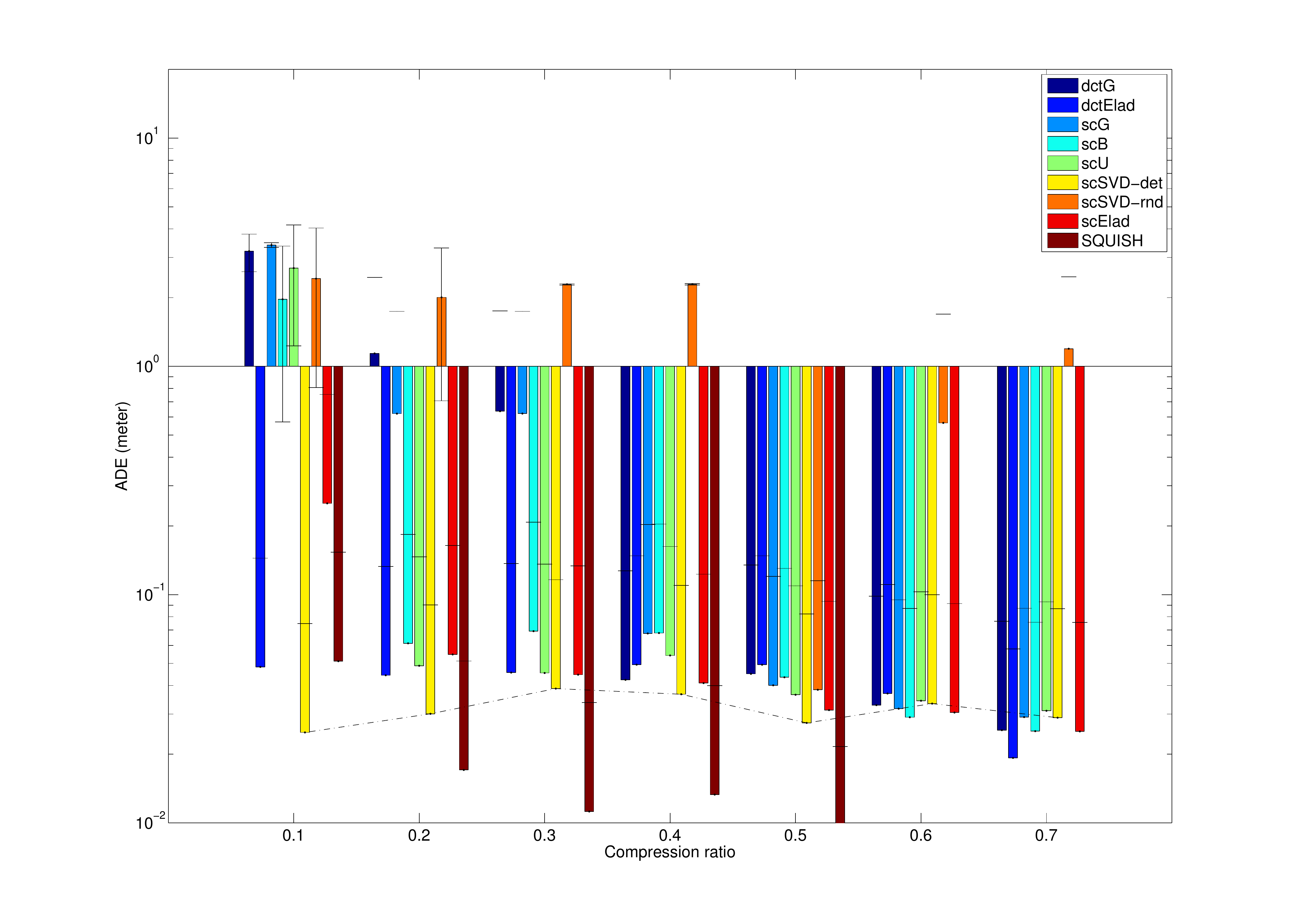}
\caption{NewYork}
\label{fig:newyork}
\end{figure*}

\begin{figure*}
\centering
\includegraphics[width=0.7\linewidth]{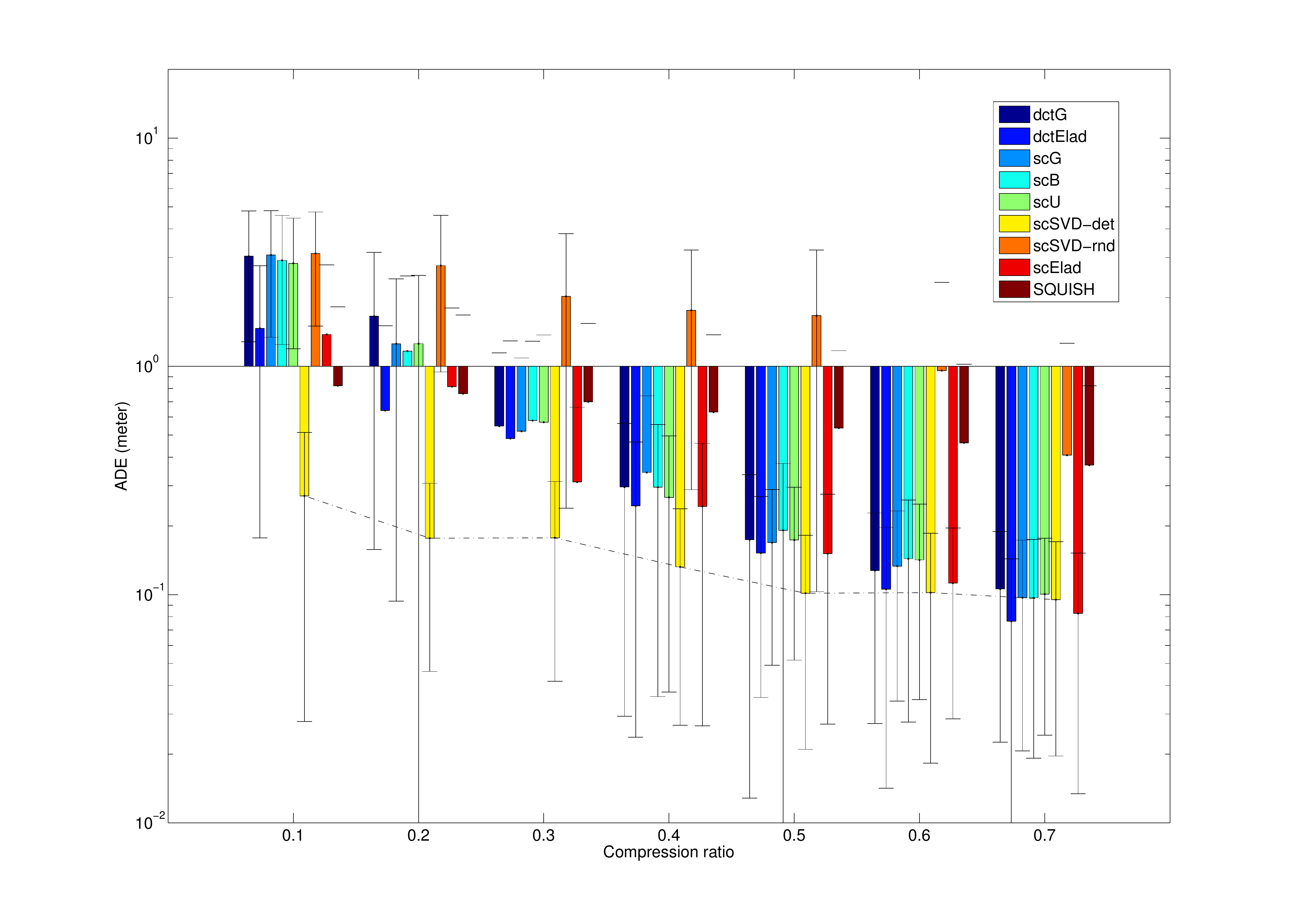}
\caption{Orlando}
\label{fig:Orlando}
\end{figure*}

\begin{figure*}
\centering
\includegraphics[width=0.7\linewidth]{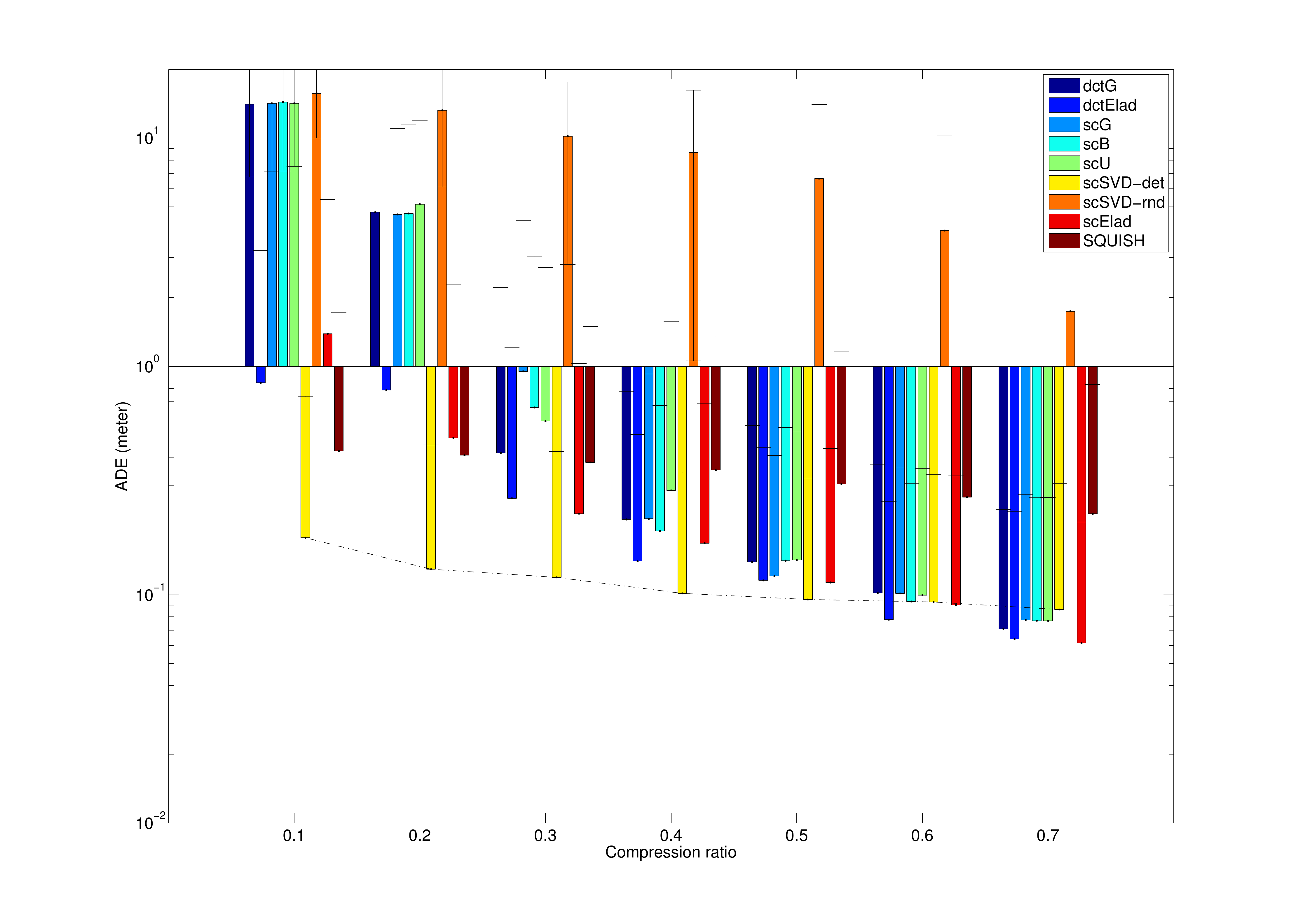}
\caption{KAIST}
\label{fig:Kaist}
\end{figure*}

\end{document}